\documentclass[a4paper,11pt]{article}
\usepackage{jheppub} 
\usepackage{lineno}
\usepackage{txfonts}
\usepackage{physics}
\usepackage{subcaption}
\usepackage{mathrsfs}
\usepackage{amsmath}
\usepackage{tikz}
\usetikzlibrary{arrows,matrix,positioning}
\usepackage{appendix}

\newcommand{{\rn}}{Reissner-Nordstr\"om}

\title{Decoherence of quantum superpositions in near-extremal Reissner-Nordstr\"om black holes with quantum gravity corrections}

\author[a,1]{Ran Li\note[1]{Corresponding author.},}

\author[a,2]{Zhong-Xiao Man\note[2]{Corresponding author.},}

\author[b,3]{Jin Wang\note[3]{Corresponding author.}}

\affiliation[a]{Department of Physics, Qufu Normal University, Qufu, Shandong 273165, China}

\affiliation[b]{Department of Chemistry, Physics and Astronomy, Stony Brook University, Stony Brook, NY 11794, USA,}

\emailAdd{liran@qfnu.edu.cn}
\emailAdd{zxman@qfnu.edu.cn}
\emailAdd{jin.wang.1@stonybrook.edu}

\abstract{We study the quantum gravity corrected decoherence of quantum superpositions in the near-extremal {\rn} black holes. By employing the effective field theory approach, we model the black hole as a quantum system coupled to an external source via a scalar field, and derive the relation between the decoherence rate and the two-point correlation function of the operators acting on the black quantum system. By utilizing the low-energy Schwarzian effective theory, which captures the boundary dynamics of the $AdS_2$ near-horizon geometry of the near-extremal {\rn} black holes, we compute the decoherence rate both in the microcanonical and canonical ensembles. We find that in the microcanonical ensemble, where the black hole energy is fixed, quantum gravity corrections do not modify the decoherence rate compared to the semiclassical prediction. However, in the canonical ensemble, where the black hole is in a thermal equilibrium state, quantum gravitational effects significantly enhance the decoherence rate at low temperatures. Our results demonstrate that even in the near-extremal limit where Hawking radiation is suppressed, quantum gravitational fluctuations can strongly influence the coherence of nearby quantum systems. }

\begin{document}

\maketitle

\section{Introduction}
\label{sec:intro}

The interplay between quantum mechanics and general relativity is one of the most profound frontiers in modern fundamental physics. A particularly interesting question concerns how quantum coherence is affected by strong gravitational fields. Black holes, with their event horizons and extreme gravitational environments, offer an unique setting where quantum and gravitational effects intertwine in a remarkable way.

Recently, Danielson, Satishchandran, and Wald (DSW) have initiated a gedanken experiment to study the decoherence of quantum spatial superposition state of a charged massive particle in the black hole spacetime \cite{Danielson:2022tdw,Danielson:2022sga}. It is demonstrated that the presence of black hole Killing horizon can inevitably decohere the quantum superposition. When the particle is placed in a spatial superposition, the associated electromagnetic or gravitational field is likewise placed in a superposition. As a result, the quantum state of the background field is also superposed. This configuration can be shown to induce a flux of soft radiation (photons or gravitons) across a Killing horizon. These soft quanta, encoding ``which-way" information about the superposition, can lead to the decoherence of the original quantum system.

The DSW decoherence effect has attracted significant attention in recent studies. In Ref. \cite{Gralla:2023oya}, the precise decoherence rate was derived for a charged massive particle in the background of a rotating Kerr black hole. To better understand its thermal origin, the DSW effect was analyzed from the perspectives of Unruh–DeWitt detector near a Rindler horizon \cite{Wilson-Gerow:2024ljx} and effective theory by modeling black hole as a quantum system at finite temperature \cite{Biggs:2024dgp}. Building on these insights, a local description of decoherence was developed in  \cite{Danielson:2024yru} by establishing the direct connection between the decoherence rate and the two-point correlation function. In our previous work \cite{Li:2024guo}, we demonstrated that quantum superpositions of charged bodies decohere near Reissner–Nordström black holes, with the effect suppressed in the extremal limit. Additionally, we computed the exact decoherence rates in de Sitter spacetime for scalar, electromagnetic, and gravitational fields \cite{Li:2024lfv}. These studies have shown that the superposition states in spacetimes with Killing horizons decohere inevitably over sufficiently long timescales. However, it has been demonstrated that decoherence can be controlled and minimized for finite durations through an optimal protocol \cite{Danielson:2025iud}. More recently, it is argued that the DSW effect allows quantum information to be teleported into a black hole at arbitrarily low energy cost, with no violation of unitarity or the generalized second law, as long as no net information is erased \cite{Kudler-Flam:2025yur}.

These studies manifest that the DSW decoherence effect can be interpreted as a form of environment-induced decoherence, which has been extensively studied \cite{Zurek:2003zz,Schlosshauer:2003zy}. While creating spatial superposition states for macroscopic objects remains experimentally challenging, notable progress has been made in platforms such as optomechanical systems \cite{Carlesso:2019cuh} and Bose–Einstein condensates \cite{Howl:2020isj}. Moreover, decoherence is deeply connected to the quantum measurement problem, which makes it play an essential role in foundational interpretations of quantum mechanics. Therefore, a deeper understanding of the DSW effect may provide new insights into the quantum nature of black holes and the black hole information problem.

It should be emphasized that previous studies of the DSW effect have been conducted primarily within a semiclassical framework, in which the radiation fields are treated quantum mechanically while the black hole background is regarded as a classical object. However, for the near extremal black hole, where the black hole temperature approaches zero, the classical descriptions break down and the quantum gravitational effects become significant \cite{Iliesiu:2020qvm}. In particular, the near-extremal {\rn} black holes exhibit an emergent near-horizon geometry of the form $AdS_2\times S_2$, which enables a low-energy effective description in terms of Jackiw–Teitelboim (JT) gravity governed by the Schwarzian action \cite{Iliesiu:2022onk}. This framework has led to some new understanding of the thermodynamics, Hawking evaporation, and absorption cross-sections of the near-extremal {\rn} black holes \cite{Kolanowski:2024zrq,Brown:2024ajk,Maulik:2025hax,Emparan:2025sao,Biggs:2025nzs}. Therefore, it is expected that this effective low-energy quantum gravitational description of near-extremal black holes may introduce corrections to the DSW decoherence effect \cite{Biggs:2025nzs}. In this way, the black hole and the sourced particle can be modeled as quantum systems simultaneously, which brings the original setup closer to a fully quantum mechanical description.

In this work, we investigate the DSW decoherence effect in the near-extremal {\rn} black hole by modeling the black hole as a quantum system. The black quantum system is assumed to be coupled with an external source particle via a scalar field. The scalar field mediates interactions between the black hole and a source particle prepared in a spatial superposition. We derive a general expression for the decoherence rate in terms of the two-point correlation function of operators localized at the boundary of $AdS_2\times S_2$, and evaluate this expression in two distinct statistical ensembles. In the microcanonical ensemble, where the total energy of the black hole is held fixed, we find that the leading behavior of the decoherence rate remains unchanged from the semiclassical prediction. Quantum gravity corrections, in this setting, do not significantly affect the infrared structure of the correlators. In contrast, in the canonical ensemble, where thermal fluctuations are included, the decoherence rate is remarkably enhanced at low temperatures due to quantum fluctuations of the boundary mode. This enhancement suggests that even as the black hole temperature approaches zero, the system retains its ability to decohere external quantum superpositions—a feature not captured by purely semiclassical analysis.

This paper is organized as follows. In Sec.\ref{sec:low_eff_nebh}, we review the thermodynamics of near-extremal {\rn} black holes and their low-energy near-horizon dynamics as described by the Schwarzian theory. In Sec.\ref{sec:Dec_BH}, we develop an effective field theory model in which the black hole is treated as a quantum system interacting with a source particle via a scalar field, and we derive an expression for the decoherence rate in terms of a two-point correlation function. In Sec.\ref{sec:Dec_RN}, we compute the decoherence rate in both the microcanonical and canonical ensembles. The final section summarizes the main findings, discusses their implications for black hole quantum dynamics, and outlines directions for future research.

\section{Low energy effective theory of the near-extremal {\rn} black holes}
\label{sec:low_eff_nebh}

In this section, we review the geometry of the near-extremal {\rn} black holes and discuss how the low energy effective dynamics near the horizon of the near-extremal {\rn} black holes can be equivalently described by a Schwarzian theory. 

\subsection{{\rn} black holes} 

The metric and gauge field of the {\rn} black hole with mass $M$ and electric charge $Q$ are given by 
\begin{eqnarray}
    ds^2&=&-f(r) dt^2+f(r)^{-1} dr^2+r^2\left(d\theta^2+\sin^2\theta d\phi^2\right)\;,\\
    A&=&-\frac{Q}{r} dt\;,
\end{eqnarray}
with
\begin{eqnarray}
    f(r)=1-\frac{2M}{r}+\frac{Q^2}{r^2}\;.
\end{eqnarray}
Here, $M$ and $Q$ are the mass and the charge of the {\rn} black hole. The black hole has two horizons, the event and the Cauchy horizons, which can be determined by the equation $f(r)=0$. They are given by
\begin{eqnarray}
    r_{\pm}=M\pm \sqrt{M^2-Q^2}\;.
\end{eqnarray}
The Hawking temperature and the classical Bekenstein-Hawking entropy of the event horizon are given by 
\begin{eqnarray}\label{Hawking_tem}
    T_H&=&\frac{r_+ - r_-}{4\pi r_+^2}\;,\\
    S&=&\pi r_+^2\;.
\end{eqnarray}

In the extremal limit, the Hawking temperature $T_H$ of the black hole vanishes, and the two horizons $r_\pm$ coincide. An extremal {\rn} black hole has a mass given by
\begin{eqnarray}
    M_0=Q\;.
\end{eqnarray}
The extremal mass $M_0$ represents the minimal mass allowed within the solution space for a given charge $Q$, ensuring that the solution describes a black hole rather than a naked singularity, in accordance with the cosmic censorship conjecture.

\subsection{Low-teperature expansion} 

We now proceed to study the low-temperature expansion of the thermodynamic quantities. Expanding around their extremal values allows us to determine the effective temperature scale at which the semiclassical approximation is expected to break down \cite{Preskill:1991tb,Maldacena:1998uz,Page:2000dk}.

The near-extremal expansion depends on the choice of ensemble. In this work, we mainly focus on the canonical ensemble, where the electric charge $Q$ of the black hole is held fixed. At small but non-vanishing temperature, one can get the expansion of the mass from Eq.\eqref{Hawking_tem} as follows 
\begin{eqnarray}
    M=r_0+2\pi^2 r_0^3 T_H^2+16\pi^3 r_0^4 T_H^3+ \mathcal{O}(T_H^4)\;,\label{M_exp}
\end{eqnarray}
where $r_0=M_0=Q$ is the radius of the extremal horizon. From this expansion, one can directly get the expansions of the horizon radii as
\begin{eqnarray}
    r_{+}&=&r_0 + 2\pi r_0^2 T_H + 10\pi^2 r_0^3 T_H^2 + \mathcal{O}(T_H^3)\;,
    \\
    r_{-}&=&r_0 - 2\pi r_0^2 T_H - 6\pi^2 r_0^3 T_H^2 + \mathcal{O}(T_H^3)\;.
\end{eqnarray}
The Bekenstein-Hawking entropy can also be expanded as 
\begin{eqnarray}
    S=S_0+4\pi^2 r_0^3 T_H+ \mathcal{O}(T_H^2)\;,
\end{eqnarray}
where $S_0=\pi r_0^2$ is the extremal black hole entropy.

The thermodynamic description of a near extremal {\rn} black hole breaks down at temperatures $T\leq E_{brk}$, where the energy scale $E_{brk}$ is given by \cite{Iliesiu:2020qvm,Kapec:2023ruw}
\begin{eqnarray}
    E_{brk}=\frac{1}{r_0^3}=\frac{1}{Q^3}\;. 
\end{eqnarray}
At low temperatures, quantum effects become significant. The near extremal {\rn} black holes can be equivalently described by a one-dimensional Schwarzian theory living on the $AdS_2$ throat \cite{Sachdev:2019bjn}.

\subsection{low-energy effective theory} 
We now consider the near-extremal geometry of the {\rn} black hole. When the black hole is close to extremality, the near-horizon geometry can be approximated by the following metric
\begin{eqnarray}
    ds^2=-\frac{\rho^2-4\pi^2r_0^4 T_H^2}{r_0^2}dt^2+\frac{r_0^2}{\rho^2-4\pi^2r_0^4 T_H^2}d\rho^2 + (\rho+r_0)^2 \left(d\theta^2+\sin^2\theta d\phi^2\right)\;,
\end{eqnarray}
where $\rho=r-r_0$. Here, the terms with $T_H^2$ represent the finite temperature correction to the near-extremal geometry. However, these corrections are not relevant to our discussion. By neglecting these corrections, one can see that the near-extremal geometry of the {\rn} black hole is exactly $AdS_2\times S^2$ geometry with both $AdS_2$ and $S^2$ radii as $r_0$. 

By applying a dimensional reduction to the near-extramal geometry, the resulting theory on $AdS_2$ can be properly matched with Jackiw-Teitelboim (JT) gravity with a negative cosmological constant \cite{Sachdev:2019bjn,Almheiri:2014cka,Jackiw:1984je,Teitelboim:1983ux}. Since there is no bulk dynamics in $AdS_2$ geometry, the entire dynamics comes from the $AdS_2$ boundary. The boundary of $AdS_2$ allows for nontrivial fluctuations corresponding to time reparametrizations. These fluctuations are the only low-energy dynamical degrees of freedom and are governed by the Schwarzian action. On the other hand, the Sachdev-Ye-Kitaev (SYK) model \cite{Sachdev:1992fk,2015escq.progE...2K} in the strong coupling limit is equivalent to the Schwarzian theory. This correspondence reveals a deep relationship in lower dimensional holographic duality \cite{Maldacena:2016hyu,Maldacena:2016upp}.

With this in mind, the Schwarzian theory captures the leading low-energy dynamics near extremality, including the deviations from extremal entropy and the specific heat at small temperatures. It plays a central role in understanding the quantum gravity corrections of near-extremal black holes. For the one-dimensional Schwarzian theory, one can perform the path integral exactly to obtain the partition function in the canonical ensemble, which is given by \cite{Mertens:2019tcm}
\begin{eqnarray}\label{partition_func}
    \mathcal{Z}(\beta,Q)=\frac{1}{4\pi^2} \left(\frac{2\pi}{\beta E_{brk}} \right)^{\frac{3}{2}} e^{S_0-\beta Q + \frac{2\pi^2}{\beta E_{brk}}}\;.  
\end{eqnarray}
Here the prefactor comes from the gravitational one-loop correction from the JT mode which dominates at low temperatures $\beta E_{brk}\gg 1$. The exponential contains the extremal entropy, extremal mass, and leading semiclassical correction to the entropy and mass terms away from extremality. 

From the partition function \eqref{partition_func}, one can obtain the density of states through an inverse Laplace transform. It can be shown that the partition function can be rewritten as
\begin{eqnarray}
    \mathcal{Z}(\beta,Q)=\int_{Q}^{+\infty} dM \frac{ e^{S_0}}{2\pi^2 E_{brk}} \sinh(2\pi \sqrt{\frac{2\left(M-Q\right)}{E_{brk}}} ) e^{-\beta M}\;.  
\end{eqnarray}
By defining the energy above the extremality bound as $E=M-Q$, the above equation can be rewritten as 
\begin{eqnarray}
    \mathcal{Z}(\beta,Q)=\int_{0}^{+\infty} dE \frac{ e^{S_0}}{2\pi^2 E_{brk}} \sinh(2\pi \sqrt{\frac{2E}{E_{brk}}} ) e^{-\beta (E+Q)}\;, 
\end{eqnarray}
from which the state density for the black hole with fixed charge $Q$ and energy $E$ can be obtained as
\begin{eqnarray}
    \rho(E)=\frac{1}{2\pi^2 E_{brk}} e^{S_0} \sinh\left(2\pi \sqrt{\frac{2E}{E_{brk}}}\right) \Theta(E)\;,
\end{eqnarray}
where $\Theta(E)$ is the Heaviside step function.

For latter calculation of quantum corrections to decoherence rate, we also need two point correlation function in Schwarzian theory. The two point correlation function of operator $\mathcal{O}$ with the conformal dimension $\Delta$ is given by \cite{Mertens:2019tcm} 
\begin{eqnarray}
    \left|\langle E_2|\mathcal{O}|E_1 \rangle \right|^2
    =\frac{2e^{-S_0}\Gamma\left(\Delta\pm i\sqrt{2E_1/E_{brk}}\pm i\sqrt{2E_2/E_{brk}}\right)}{\left(2E_{brk}^{-1}\right)^{2\Delta}\Gamma(2\Delta)}\;.
\end{eqnarray}
Here, we are using a standard convention where there is an implicit product over all choices of sign appearing in gamma functions. Then for a black hole with fixed charge, the spectral density for the two-point correlation function with quantum gravity-corrections is given by \cite{Biggs:2025nzs}
\begin{eqnarray}\label{spectral_density}
    \int dt e^{-i\omega t} \langle E|O(t) O(0)|E\rangle=2\pi \rho(E-\omega)  \left|\langle E|\mathcal{O}|E-\omega \rangle \right|^2\;.
\end{eqnarray}
Here, the calculation of two-point correlation function below the energy scale $E_{brk}$ involves integrating over the Schwarzian modes.

\section{Decoherence of quantum superposition in black hole spacetime}
\label{sec:Dec_BH}

In this section, we consider the quantum superposition experiment performed in the black hole spacetime from the effective field theory approach. We will try to derive the relation between the decoherence rate and the two-point correlation function.

\subsection{Effective field theory approach of DSW experiment} 

As discussed in the introduction, the DSW gardenken experiment of quantum superposition can also be realized by treating the black hole as a black quantum system in the effective field theory approach. The temperature of the black quantum system matches the Hawking temperature of the corresponding black hole.

For this purpose, we consider the dephasing channel, which is a well known decoherence channel in open quantum system. The black hole is modeled as a quantum system localized at a point in Minkowski spacetime. The decoherence effect can be understood as the black quantum system $B$ absorbing the soft radiation emitted by a charged "massive" particle $A$ prepared in a spatial superposition state. Without loss of generality, we assume that the black quantum system $B$ is located at the origin of Minkovski spacetime and the particle $A$ is far from the black quantum system $B$.  

Although the original DSW decoherence mechanism arises from the interaction between the black hole and electromagnetic or gravitational fields, it has been shown that a scalar field model can also capture the essential features of this effect. Therefore, we introduce the following action for the entire system \cite{Biggs:2024dgp}
\begin{eqnarray}\label{action}
    S&=&S_{B}-\frac{1}{2}\int d^4x \left(\nabla \phi\right)^2 + \int d^4x \lambda(t) \phi(t,\vec{x}) \sigma_3 \delta^{(3)}(\vec{x}-\vec{x}_{A})
    \nonumber\\&&+ g \int d^4 x O(t) \phi(t,\vec{x})  \delta^{(3)}(\vec{x}-\vec{x}_{B})\;.
\end{eqnarray}
Here, $S_{B}$ denotes the action of the black quantum system located at the origin of Minkovski spacetime. The second term denotes the action of free scalar field. The third denotes the interaction between the particle $A$ and the scalar field with $\lambda(t)$ being the time dependent coupling. The operator $O(t)$ can be treated as the response of the external perturbation $\phi$. Therefore, $O(t)$ acts on the black quantum system and the forth term is the interaction between the black quantum system and the scalar field with the coupling constant given by $g$. We don't consider the dynamics of the soured particle $A$. In this action, the interaction between the black quantum system and the particle $A$ is mediated by the scalar field. Note that the coupling $\lambda(t)$ will be specified in the following and the coupling $g$ can be fixed in the next section.

A convenient way to proceed is to integrate out the scalar field and only keep the interaction between the black quantum system $B$ and the particle source $A$. For this aim, we consider the scalar equation with the source term as follows 
\begin{eqnarray}
    \Box \phi(t,\vec{x}) = -\lambda(t) \sigma_3 \delta^{(3)}(\vec{x}-\vec{x}_{A})- g  O(t)  \delta^{(3)}(\vec{x}-\vec{x}_{B})\;,
\end{eqnarray}
Using the Green's function, the formal solution to the scalar equation can be written in the form of 
\begin{eqnarray}\label{scalr_sol}
    \phi(t,\vec{x})=- \int d^4 x G(t-t',\vec{x}-\vec{x}') \left(\lambda(t) \sigma_3 \delta^{(3)}(\vec{x}-\vec{x}_{A})+ g  O(t)  \delta^{(3)}(\vec{x}-\vec{x}_{B})\right)\;,
\end{eqnarray}
where $G(t-t',\vec{x}-\vec{x}')$ is the Green's function.

Here, we consider the interaction between the source particle $A$ and the black quantum system $B$ is instantaneous, which means that the Green's function is of the form 
\begin{eqnarray}\label{Green_fun}
    G(t-t',\vec{x}-\vec{x}')= \delta(t-t')\cdot \frac{1}{4\pi |\vec{x}-\vec{x}'|}\;,
\end{eqnarray}
where the delta function denotes the interaction is instantaneous and $|\vec{x}-\vec{x}'|$ denotes the spatial distance between the two subsystems \cite{Biggs:2024dgp}.

By substituting the formal solution \eqref{scalr_sol} of scalar field into the last term of the total action \eqref{action}, and using the explicit form \eqref{Green_fun} of the Green's function, one can get the interaction between the black quantum system $B$ and the source particle $A$ as 
\begin{eqnarray}\label{S_int}
    S_{int} = -\frac{g}{4\pi d} \int dt \lambda(t) O(t) \sigma_3 \;,
\end{eqnarray}
where $d$ is the spatial distance between the source particle $A$ and the black quantum system $B$. This term also determines the interaction Hamiltonian of the source particle and the black quantum system.

\subsection{Decoherence rate of quantum superposition}

We now consider the time evolution of the system combined by the source particle $A$ and the black quantum system $B$. We treat the subsystem $A$ as the central system, while treating $B$ as the environment. The theory of open quantum system can be now employed to study the decoherence effect. 

We assume that the time dependent coupling $\lambda(t)$ is a function as \cite{Danielson:2022sga}
\begin{eqnarray}
    \lambda(t)=\begin{cases}
\lambda_0, & |t|<T/2\;,\\
0, & t<-T/2-T_1 \;\;\textrm{or}\;\; t>T/2+T_2\;,
\end{cases}
\end{eqnarray}
where $T$ is the time that the experimenter holds the superposition state of the subsystem $A$ and $T_1$ and $T_2$ are the time that the times that used to separate and recombine the superposition. One can also treat the time $T$ as the time that the coupling between the source particle and the black quantum system is switched on \cite{Kudler-Flam:2025yur}.

For a superposition experiment, we consider the source particle $A$ is initially in the state described by the density matrix as 
\begin{eqnarray}
    \hat{\rho}_A = \left[\begin{matrix}
\rho_{11} & \rho_{12} \\
\rho_{21} & \rho_{22}
\end{matrix}\right]\;.
\end{eqnarray}
For such a density matrix, it is well known that the off diagonal terms represent the coherence of the quantum state. Therefore, we mainly focus on the time evolution of the off diagonal terms and omit the dynamics of the source particle itself.

The initial state of the total system is given by 
\begin{eqnarray}
    \hat{\rho}_{in}=\hat{\rho}_A\otimes \hat{\rho}_B=\hat{\rho}_A\otimes |0\rangle\langle 0|\;,
\end{eqnarray}
where we have assumed that the initial state of the black quantum system is in the vacuum state. 

In the interaction picture, the time evolution of the density matrix is governed by the von Neumann equation \cite{breuer2002theory}
\begin{eqnarray}
    \frac{d\hat{\rho}(t)}{dt}=-i\left[\hat{H}_{int},\hat{\rho}(t)\right]\;,
\end{eqnarray}
for which the formal solution is given by 
\begin{eqnarray}\label{rho_form}
    \hat{\rho}(t) =\hat{U}(t) \hat{\rho}_{in} \hat{U}(t)^\dagger\;.
\end{eqnarray}
For the interaction $S_{int}$ in Eq.\eqref{S_int}, the unitary evolution operator can be given by 
\begin{eqnarray}
    \hat{U}=\mathcal{T} \left\{\textrm{exp}\left[i \int_{-\infty}^t d\tau \hat{\lambda}(\tau) \hat{O}(\tau) \sigma_3 \right] \right\}\;,
\end{eqnarray}
where $\mathcal{T}$ denotes the temporal order product operator and we have introduced $\hat{\lambda}(t)=\frac{g}{4\pi d}\lambda(t)$ for convenience. Here, we have assumed that the interaction Hamiltonian is commutative with the free part of the Hamiltonian. Note that the coupling is only opened in a finite time, it is reasonable to take the integral of time $t$ from $-\infty$.

The Dyson expansion for the evolution operator can be written as 
\begin{eqnarray}
    \hat{U}=I+i \sigma_3 \int_{-\infty}^t d\tau \hat{\lambda}(\tau) \hat{O}(\tau) - \frac{1}{2} \int_{-\infty}^t d\tau_1 \int_{-\infty}^t d\tau_2 \mathcal{T}\left[\hat{\lambda}(\tau_1) \hat{O}(\tau_1)\hat{\lambda}(\tau_2) \hat{O}(\tau_2)\right]+\cdots\;,
\end{eqnarray}
where the high order expansions are omitted here. 

By substituting this expansion into Eq.\eqref{rho_form}, one can get the density matrix of the total system at time $t$ as
\begin{eqnarray}
    \hat{\rho}(t)&=& \hat{\rho}_{in} +i \sigma_3 \int_{-\infty}^t d\tau \hat{\lambda}(\tau) \left[\hat{O}(\tau),\hat{\rho}_{in}\right] 
    +\int_{-\infty}^t d\tau_1 \int_{-\infty}^t d\tau_2 \hat{\lambda}(\tau_1) \hat{O}(\tau_1)\sigma_3 \hat{\rho}_{in}\sigma_3 \hat{\lambda}(\tau_2) \hat{O}(\tau_2)\nonumber\\
    &&-\frac{1}{2} \int_{-\infty}^t d\tau_1 \int_{-\infty}^t d\tau_2 \left\{ \hat{\lambda}(\tau_1) \hat{O}(\tau_1)\hat{\lambda}(\tau_2) \hat{O}(\tau_2), \hat{\rho}_{in} \right\}\;,
\end{eqnarray}
where $\{*,*\}$ denotes the anti-commutator. 

In order to obtain the reduced density matrix of the source particle $A$, one has to take the partial trace over the black quantum system $B$. Considering the form of the interaction of these two subsystems, the $n$-th order term in the reduced density matrix comes from the $n$-point correlation function of the operator $\hat{O}$ that is acting on the black quantum system. We assume that for the free theory of the total system, the $n$-point correlation function vanishes for odd $n$.

Taking the partial trace over the black quantum system, one can get the reduced density matrix of the particle $A$ at the time that the coupling is switched off as 
\begin{eqnarray}
    \hat{\rho}_A(+\infty)&=&\hat{\rho}_A+\frac{1}{4}\left(\sigma_3 \hat{\rho}_A \sigma_3 - \hat{\rho}_A\right) \mathcal{N}
    \nonumber\\
    &=&\hat{\rho}_A-\frac{1}{2}\left[\sigma_3,\left[\sigma_3,\hat{\rho}_A\right] \right] \mathcal{N}
    \;,
\end{eqnarray}
where 
\begin{eqnarray}\label{N_eq}
    \mathcal{N} &=& 4 \int_{-\infty}^{+\infty} d\tau_1 \int_{-\infty}^{+\infty} d\tau_2  \hat{\lambda}(\tau_1) \hat{\lambda}(\tau_2) \left\langle\hat{O}(\tau_1)\hat{O}(\tau_2)\right\rangle \;\nonumber\\
    &=&\left(\frac{g}{2\pi d}\right)^2 \int_{-\infty}^{+\infty} d\tau_1 \int_{-\infty}^{+\infty} d\tau_2  \lambda(\tau_1) \lambda(\tau_2) \left\langle\hat{O}(\tau_1)\hat{O}(\tau_2)\right\rangle \;.
\end{eqnarray}
Note that in the second line, we have restored the definition of $\hat{\lambda}$. 

It is clear that the reduced density matrix of the source particle after the experiment has the explicit form of 
\begin{eqnarray}
    \hat{\rho}_A(t)=  \left[\begin{matrix}
\rho_{11} & \left(1-\frac{1}{2}\mathcal{N}\right)\rho_{12} \\
\left(1-\frac{1}{2}\mathcal{N}\right)\rho_{21} & \rho_{22}
\end{matrix}\right]\;.
\end{eqnarray}
Here, $\mathcal{N}$ describes the decoherence in the superposition state of the source particle $A$. However, this is a perturbative result. $\mathcal{N}$ is also relevant to the particle number that is absorbed by black hole horizon. As such, $\mathcal{N}$ is named as the decoherence function.

However, there is a subtlety in the above derivation. The Dyson expansion is only valid when $\hat{\lambda}(\tau) \hat{O}(\tau)t$ is small enough. Otherwise, the high-order contributions cannot be neglected. This is analogous to the case that the perturbation approach of the dynamics of Unruh-DeWitt detector \cite{Louko:2007mu,Han:2025jcu}. A more rigorous derivation of the decoherence was presented in \cite{Wilson-Gerow:2024ljx}, where Schwinger-Keldysh path integral formalism \cite{schwinger1961brownian,keldysh1965diagram} (see \cite{sieberer2016keldysh} for a nice review on the topic) was utilized to determine the reduced density matrix at late time. It is shown that the reduced density matrix of the source particle $A$ is given by 
\begin{eqnarray}
    \hat{\rho}_A(t)=  \left[\begin{matrix}
\rho_{11} & \mathcal{F}_{12}\rho_{12} \\
\mathcal{F}_{21}\rho_{21} & \rho_{22}
\end{matrix}\right]\;,
\end{eqnarray}
where $\mathcal{F}$ is the influence functional. It is clear that the magnitude of the influence functional is relevant to quantify the decoherence. The magnitude of the influence functional can be written as 
\begin{eqnarray}
    |\mathcal{F}|=\textrm{exp}\left(-\frac{1}{2}\mathcal{N}\right)\;,
\end{eqnarray}
where $\mathcal{N}$ is the decoherence function given by Eq.\eqref{N_eq}.

We now estimate that the decoherence function $\mathcal{N}$ is proportional to the time $T$ that the coupling is switched on. One can roughly approximate the function $\lambda(t)$ as a rectangular wave function. By using the Fourier transformation 
\begin{eqnarray}
    \lambda(t)=\int \frac{d\omega}{2\pi} \tilde{\lambda}(\omega) e^{-i\omega t}\;,
\end{eqnarray}
one can get 
\begin{eqnarray}
    \mathcal{N} &\approx& \left(\frac{g}{2\pi d}\right)^2 \int \frac{d\omega}{2\pi}  \tilde{\lambda}(\omega)\tilde{\lambda}^*(\omega) \mathcal{S}(\omega)\;,
\end{eqnarray}
where the Fourier transform $\mathcal{S}(\omega)$  of the correlation function is given by 
\begin{eqnarray}
    \mathcal{S}(\omega)&=&\int d(\tau_1-\tau_2) e^{i\omega(\tau_1-\tau_2)} \left\langle\hat{O}(\tau_1)\hat{O}(\tau_2)\right\rangle\\
    &=&\int dt e^{i\omega t} \left\langle\hat{O}(t)\hat{O}(0)\right\rangle\;.
\end{eqnarray}
Here we assume the time translation symmetry of the correlation function.

Since $\lambda(t)$ can be approximated as a rectangular wave function, for large $T$, $\tilde{\lambda}(\omega)$ is highly band limited near $\omega\sim 0$. Therefore, we have
\begin{eqnarray}
   \mathcal{N} &\approx& \left(\frac{g}{2\pi d}\right)^2\mathcal{S}(\omega=0) \int \frac{d\omega}{2\pi}  \tilde{\lambda}(\omega)\tilde{\lambda}^*(\omega)\nonumber\\
    &\approx&  \left(\frac{g}{2\pi d}\right)^2\mathcal{S}(\omega=0)  \int dt \lambda^2(t) \nonumber\\ 
    &\approx& T \left(\frac{g \lambda_0}{2\pi d }\right)^2  \mathcal{S}(\omega=0)\;.
\end{eqnarray}
Note that, in the second line, we have used the Parseval theorem to convert the integral in the frequency domain to the time domain.

It is clear that the decoherence function $\mathcal{N}$ is proportional to the time $T$ that the coupling is switched on. On the other hand, the decoherence function $\mathcal{N}$ is related to the decoherence rate by 
\begin{eqnarray}
    \frac{\mathcal{N}}{4}=\Gamma T\;.
\end{eqnarray}
In this respect, for the scalar monopole coupling interaction given in Eq.\eqref{S_int}, one can get the decoherence rate as
\begin{eqnarray}
    \Gamma=\left(\frac{g\lambda_0}{4\pi b}\right)^2  \mathcal{S}(\omega=0)\;.
\end{eqnarray}
This equation gives us the relation between the decoherence rate of quantum superposition state and the two-point correlation function of the black quantum system. We are now in the position to evaluate the zero frequency limit of the Fourier transformation of the two-point correlation function of the black quantum system.

\section{Quantum gravity corrected decoherence rate of near-extremal {\rn} black hole} 
\label{sec:Dec_RN}

In this section, we discuss the quantum gravity corrected decoherence rate for the near-extremal {\rn} black hole.

\subsection{The coupling constant}
We now consider how to fix the coupling constant between the black quantum system and the source particle \cite{Biggs:2025nzs}. In the semiclassical regime, the two-point correlation function for the scalar operator $(\Delta=1)$ is totally determined by the conformal symmetry, which can be given by 
\begin{eqnarray}\label{S_semi}
    S(\omega)= \frac{2\pi \omega}{1-e^{-\beta\omega}}\;,
\end{eqnarray}
where $\beta$ is the inverse Hawking temperature. At the zero frequency limit, one has 
\begin{eqnarray}
   S(\omega=0)\simeq \frac{2\pi }{\beta}\;,
\end{eqnarray}
which gives an expression of the semiclassical decoherence rate as \cite{Li:2024guo}
\begin{eqnarray}\label{Gamma_g}
    \Gamma=\left(\frac{g\lambda_0}{4\pi b}\right)^2 \frac{2\pi }{\beta}\;.
\end{eqnarray}

On the other hand, at the low energy limit $\omega\beta\ll 1$, the decoherence rate can be related to the semiclassical absorption section $\sigma(\omega)$ as \cite{Biggs:2024dgp}
\begin{eqnarray}\label{Gamma_s}
    \Gamma_{\textrm{semi}}= \left(\frac{\lambda_0}{4\pi b}\right)^2 \left.\frac{2\omega\sigma(\omega)}{e^{\beta\omega}-1}\right|_{\omega\rightarrow 0}\nonumber\\
    =\left(\frac{\lambda_0}{4\pi b}\right)^2 \frac{2}{\beta} \sigma(\omega=0)
    \;.
\end{eqnarray}
It is known that the absorption section for the spherically symmetric black hole at the low energy limit is given by its horizon area \cite{Das:1996we,Cvetic:1997ap}. Therefore, we have the semiclassical decoherence rate as 
\begin{eqnarray}\label{Gamma_s}
    \Gamma_{\textrm{semi}}= \left(\frac{\lambda_0}{4\pi b}\right)^2 \frac{8\pi r_+^2}{\beta}\;.
\end{eqnarray}
This result is also consistent with the result in \cite{Gralla:2023oya} by taking the zero rotating limit. In the extremal limit when $T_H\rightarrow 0$, the decoherence rate $\Gamma\rightarrow 0$. This means that the coherence of the quantum superposition state in the extremal {\rn} black hole can be maintained for very long time \cite{Li:2024guo}.

By matching the expressions Eq.\eqref{Gamma_g} and Eq.\eqref{Gamma_s} of the decoherence rate at the semiclassical regime, we can fix the coupling constant as 
\begin{eqnarray}
    g=2r_+\;.
\end{eqnarray}
Therefore, the decoherence rate can be written as 
\begin{eqnarray}\label{Dec_S}
      \Gamma=\left(\frac{\lambda_0 r_+}{2\pi b}\right)^2 \mathcal{S}(\omega=0) \;.
\end{eqnarray}

\subsection{Decoherence rate in microcanonical ensemble}

We now evaluate the spectral density of the two-point correlation function in the microcanonical ensemble. For microcanonical ensemble, we consider the energy $E$ is fixed. From Eq.\eqref{spectral_density}, we have
\begin{eqnarray}
    \mathcal{S}_E(\omega)&=&\int dt e^{i\omega t} \langle E|O(t)O(0)|E\rangle\nonumber\\
    &=&2\pi \rho(E+\omega)  \left|\langle E|\mathcal{O}|E+\omega \rangle \right|^2\nonumber\\
   &=& \frac{2\pi \omega\sinh2\pi \sqrt{\frac{2(E+\omega)}{E_{brk}}} }{\cosh 2\pi \sqrt{\frac{2(E+\omega)}{E_{brk}}}-\cosh 2\pi \sqrt{\frac{2E}{E_{brk}}}}\;,
\end{eqnarray}
where $E$ is the black hole energy above the extremal value. 

In the semiclassical limit when $E_{brk}\ll E$, the Wightman function is reduced to the form of 
\begin{eqnarray}
    \mathcal{S}_{\textrm{semi}}(\omega)\simeq \frac{2\pi \omega}{1- e^{-\frac{2\pi\omega}{\sqrt{2EE_{brk}}}}}\;,
\end{eqnarray}
which coincides with Eq.\eqref{S_semi} with the temperature given by 
\begin{eqnarray}
    T=\frac{\sqrt{2EE_{brk}}}{2\pi}\;.
\end{eqnarray}
This means that in the semiclassical limit, the two-point correlation function is determined by the conformal symmetry. 

The decoherence rate is governed by $S_E(\omega=0)$. It can be shown that 
\begin{eqnarray}
    S_E(\omega=0)=\sqrt{2E E_{brk}}\;,
\end{eqnarray}
which gives the decoherence rate in the microcanonical ensemble as 
\begin{eqnarray}
    \Gamma= \left(\frac{\lambda_0}{4\pi b}\right)^2 4 r_+^2 \sqrt{2E  E_{brk}}\;.
\end{eqnarray}
This is the quantum gravity corrected decoherence rate in the microcanonical ensemble, valid in the near-extremal and low-temperature regime where semiclassical thermodynamics breaks down.

When both the quantum gravity result (via Schwarzian theory) and the semiclassical result are expressed in terms of the same physical variable-namely, fixed energy above extremality-they give exactly the same decoherence rate 
\begin{eqnarray}
    \Gamma=\Gamma_{\textrm{semi}}=\left(\frac{\lambda_0}{4\pi b}\right)^2 \frac{8\pi r_+^2}{\beta}\;.
\end{eqnarray}
This means that the quantum gravity correction doesn’t change the decoherence rate at leading order when working at fixed energy.

\subsection{Decoherence rate with quantum gravity corrections}

Now we consider the case that the black quantum system is in a finite temperature equilibrium state. We have to evaluate the thermal correlation function in the canonical ensemble. The thermal correlation function $S(\omega)$ can be written as 
\begin{eqnarray}
    S(\omega)&=&\int dt e^{i\omega t} \langle O(t) O(0)\rangle\nonumber\\
   &=&\frac{1}{\mathcal{Z}(\beta,Q)} \int dt e^{i\omega t} \int_{0}^{+\infty}dE e^{-\beta (E+Q)} \rho(E) \langle E|O(t)O(0)|E\rangle\nonumber\\
   &=&\frac{2\pi}{\mathcal{Z}(\beta,Q)} \int_{0}^{+\infty}dE e^{-\beta (E+Q)} \rho(E) \rho(E+\omega)  \left|\langle E|\mathcal{O}|E+\omega \rangle \right|^2\nonumber\\
   &=&2\beta \omega  \sqrt{\frac{\beta E_{brk}}{2\pi}} e^{-\frac{2\pi^2}{\beta E_{brk}}} \int_{0}^{+\infty} dE e^{-\beta E} \frac{\sinh2\pi \sqrt{\frac{2(E+\omega)}{E_{brk}}} \sinh 2\pi \sqrt{\frac{2E}{E_{brk}}}}{\cosh 2\pi \sqrt{\frac{2(E+\omega)}{E_{brk}}}-\cosh 2\pi \sqrt{\frac{2E}{E_{brk}}}}\;.
\end{eqnarray}
The integral is hard to evaluate. However, the decoherence is only relevant to the zero frequency limit of the Wightman function $S(\omega)$. We can first expand the integral as the series of $\omega$ and extract the zero frequency limit as 
\begin{eqnarray}
     S(\omega=0)=2 \left(\frac{\beta E_{brk}}{2\pi}\right)^{\frac{3}{2}} e^{-\frac{2\pi^2}{\beta E_{brk}}} \int_{0}^{+\infty} dE e^{-\beta E} \sqrt{\frac{2E}{E_{brk}}}
     \sinh 2\pi \sqrt{\frac{2E}{E_{brk}}}\;.
\end{eqnarray}
The integral can be obtained in a compact form as
\begin{eqnarray}
     S(\omega=0)=2 \left[\frac{1}{\beta}\left(\frac{\beta E_{brk}}{2\pi}\right)^{\frac{1}{2}}e^{-\frac{2\pi^2}{\beta E_{brk}}}+\frac{\left(4\pi^2+\beta E_{brk}\right)}{4\pi \beta} \textrm{Erf}\left(\sqrt{\frac{2\pi^2}{\beta E_{brk}}}\right)  \right]\;,
\end{eqnarray}
where $\textrm{Erf}(x)$ is the error function. Substituting this result into Eq.\eqref{Dec_S}, one can finally get the quantum gravity corrected decoherence rate in the near extremal {\rn} black hole as
\begin{eqnarray}
     \Gamma=\left(\frac{\lambda_0 }{4\pi b}\right)^2 \frac{8r_+^2}{\beta} \left[\left(\frac{\beta E_{brk}}{2\pi}\right)^{\frac{1}{2}}e^{-\frac{2\pi^2}{\beta E_{brk}}}+\frac{\left(4\pi^2+\beta E_{brk}\right)}{4\pi} \textrm{Erf}\left(\sqrt{\frac{2\pi^2}{\beta E_{brk}}}\right)  \right]\;.
\end{eqnarray}
Comparing the semiclassical result in Eq.\eqref{Gamma_s}, the quantum gravity corrections are manifested in the square brackets.

In the low temperature limit $\beta E_{brk}\gg 1$ where the quantum gravity correction dominates, we can further expand the results as 
\begin{eqnarray}
    S(\omega=0)=\frac{2}{\beta} \left(\sqrt{\frac{2\beta E_{brk}}{\pi}}+\frac{(2\pi)^{\frac{3}{2}}}{3}\frac{1}{\sqrt{\beta E_{brk}}}+\cdots\right)
\end{eqnarray}
The dominate contribution to the decoherence is given by
\begin{eqnarray}
    S(\omega=0)=\sqrt{\frac{8E_{brk}}{\pi \beta}}\;. 
\end{eqnarray}
Finally, we can get the decoherence rate of the near extremal {\rn} black hole in the low temperature limit is given by 
\begin{eqnarray}
  \Gamma=\left(\frac{\lambda_0}{4\pi b}\right)^2 \frac{8r_+^2}{\beta} \sqrt{\frac{2 \beta E_{brk}}{\pi }}\;.
\end{eqnarray}
Comparing the the semiclassical result, we have 
\begin{eqnarray}
    \frac{\Gamma}{\Gamma_{\textrm{semi}}} =  \sqrt{\frac{2 \beta E_{brk}}{\pi }} \gg 1\;.
\end{eqnarray}
Therefore, the decoherence rate of the near extremal {\rn} black hole is significantly enhanced by the quantum gravity corrections.

In the canonical ensemble (used in the main body of the paper), one integrates over a thermal distribution of energy levels with Boltzmann weight, and quantum gravity effects (via the Schwarzian partition function) introduce an enhancement in the decoherence rate at low temperature. In contrast, in the microcanonical ensemble, where the black hole energy is sharply fixed, no such enhancement arises-the Schwarzian and semiclassical results agree at leading order.

This conclusion seems different from the results obtained in \cite{Emparan:2025sao,Biggs:2025nzs} where the quantum corrected absorption cross sections for the scalar field in the near extremal {\rn} black hole is calculated. It is shown there that the absorption cross sections in both microcanonical and canonical ensemble are influenced by the Schwarzian corrections.

\section{Conclusion and discussion}
\label{sec:con_dis}

In this work, we investigated the quantum gravity-corrected decoherence of spatial quantum superpositions near near-extremal Reissner–Nordström (RN) black holes. Unlike the original semiclassical approach, the black hole is treated as a black quantum system with crucial quantum gravity corrections from the Schwarzian theory. The basic idea is that the {\rn} black holes exhibit an emergent $AdS_2\times S_2$ geometry, enabling a dimensional reduction and effective low-energy description governed by the Schwarzian action. This approach allows for exact computation of thermal two-point correlation functions, which are directly related to the decoherence rate of the source particle.

We carefully distinguish between the microcanonical ensemble, where the black hole energy is fixed, and the canonical ensemble, where the system is thermally distributed. Our results reveal a clear distinction between these two ensembles. In the microcanonical ensemble, quantum gravity corrections encoded in the Schwarzian theory do not change the semiclassical decoherence rate. In contrast, within the canonical ensemble, thermal averaging over energy levels leads to significant enhancement of the decoherence rate. This enhancement arises from the quantum fluctuations at the $AdS_2$ boundary, which become increasingly relevant as the black hole approaches extremality.

To facilitate analytic calculations, we model the interaction between the black hole and the source particle via a scalar field. While this simplifies the analysis and captures essential features of decoherence, the scalar field model should be interpreted cautiously \cite{Kudler-Flam:2025yur,Biggs:2024dgp,Li:2024lfv}. It should be noted that there are no massless scalar particles in our universe. Nevertheless, the scalar model serves as a useful proxy for gaining qualitative insights into horizon-induced decoherence effect. A more realistic treatment involving electromagnetic and gravitational radiation remains an important direction for future work.

In studying the decoherence of quantum superpositions near black holes, an important generalization may arise in black hole spacetimes with inherently non-equilibrium features, such as the black holes in de Sitter background. In de Sitter black hole spacetimes, this naturally happens because there are two horizons: the black hole horizon and the cosmological horizon. Each associates a different Hawking temperature. As a result, a particle near such a black hole is equivalent to interacting with two thermal baths with different temperatures. This creates a non-equilibrium situation, which can influence the decoherence of the quantum superposition \cite{PhysRevA.99.042320,Zhang:NJP}. To study this properly, we can use tools like the Schwinger–Keldysh formalism and the influence functional, which are designed to handle non-equilibrium quantum systems.


\acknowledgments  

Z.X.M. would like to acknowledge support from the Shandong Provincial Natural Science Foundation under projects ZR2023LLZ015 and ZR2024LLZ012.


\bibliographystyle{JHEP}
\bibliography{biblio.bib}
\end{document}